\documentclass[12pt]{appolb}
\usepackage[cp1250]{inputenc}
\usepackage[T1]{fontenc}
\usepackage{amsmath}
\usepackage[dvips]{graphicx}
\usepackage{algorithmic}
\usepackage{algorithm}
\usepackage{hyperref}
\usepackage{url}
\usepackage{calc}
\newlength{\mgrlewy}\newlength{\mgrprawy}
\newlength{\mgrgora}\newlength{\mgrdol}
\newlength{\mgrstopka}\newlength{\mgrdoglowki}
\newlength{\pomc}
\newcount\tempcntr
 \advance\pomc-.5\baselineskip
 \divide\pomc\baselineskip
 \tempcntr=\pomc
\begin{document}
\title{Automatic Trading Agent. RMT based Portfolio Theory and
Portfolio Selection
\thanks{\url{snarska@th.if.uj.edu.pl}\\ \url{jakub@krzych.art.pl}}}
\author{Małgorzata Snarska, Jakub Krzych
 \address{M. Smoluchowski Institute of Physics, Jagellonian
University, 30-059 Cracow, Reymonta 4, Poland}
 } \maketitle
\begin{abstract}\noindent
Portfolio theory is a very powerful tool in the modern investment
theory. It is helpful in estimating risk of an investor's
portfolio, which arises from our lack of information, uncertainty
and incomplete knowledge of reality, which forbids a perfect
prediction of future price changes. Despite of many advantages
this tool is not known and is not widely used among investors on
Warsaw Stock Exchange. The main reason for abandoning this method
is a high level of complexity and immense calculations. The aim of
this paper is to introduce an automatic decision - making system,
which allows a single investor to use such complex methods of
Modern Portfolio Theory (MPT). The key tool in MPT is an analysis
of an empirical covariance matrix. This matrix, obtained from
historical data is biased by such a high amount of statistical
uncertainty, that it can be seen as random. By bringing into
practice the ideas of Random Matrix Theory (RMT), the noise is
removed or significantly reduced, so the future risk and return
are better estimated and controlled. This concepts are applied to
the Warsaw Stock Exchange Simulator \url{http://gra.onet.pl}. The
result of the simulation is 18\% level of gains in comparison for
respective 10\% loss of the
Warsaw Stock Exchange main index WIG.\vspace{0.5cm} \\
\noindent \normalsize{\textbf{Keywords:} {Random Matrix Theory,
Gaussian Filtering, Portfolio Optimization}}
\end{abstract}
\fussy
\section{Portfolio theory - setting the stage}
Investments in stock securities like shares, currencies or
different types of derivatives are generally treated as very
risky. Ability to predict future movements in prices (price
changes) allows one to
minimize the risk.\\
Modern Portfolio Theory (MPT) refers to an investment strategy
that seeks to construct an optimal portfolio by considering the
relationship between risk and return. MPT suggests that the
fundamental issue of capital investment should no longer be to
pick out dominant stocks but to diversify the wealth among many
different assets. The success of investment does not purely depend
on return, but also on the risk, which has to be taken into
account. Risk itself is influenced by the correlations between
different assets, thus the portfolio selection process represents
Let us briefly remind several key tools and concepts, that MPT
uses, i.e. the Markowitz's Model, which is crucial in further
analysis.
\subsection{Elementary definitions and the Markowitz's Model} The
efficient portfolio theory was first introduced by Harry
M.Markowitz in $1952$  \cite{Marko:62}. He decided not to analyze
the return, risk and volatility of single stocks in a portfolio,
but considering portfolio (groups of shares) as a whole. In order
to manage this problem, he introduced a simple statistical measure
- correlation, which links up the changes in prices of an
individual assets with all other changes in price of assets in a
given portfolio.
\subsubsection{Construction of an efficient portfolio of multiple
assets} Consider $T$ quotations of the $i$ -th stock and introduce
a vector of returns $r_{i,1}$,where $r_{i,t}$, $t = 1,\ldots,T$ is
the observed realization of a random variable $r_i$ .Denote
$S_i(t)$ - time series of prices for a certain stock $i$. Then
\begin{equation}\label{rown:1}
 r _{i, t} = \ln S_i(t+1) -\ln S_i (t)
\end{equation}
and $\ln$ is a natural logarithm. Then the expected return of a
single asset is given by
\begin{equation}\label{rown:9}
R_i = E(r_i)= \hat{r}_i =\bar{r}_i = \frac{1}{T}\sum_{t=1}^T
r_{i,t}
\end{equation}
If additionally $N$ denotes the number of assets in a portfolio,
then $\mathbf{w}$ is a vector of weights (ratio of different
stocks in a portfolio). We have to then impose a budget constraint
\begin{equation}\label{rown:11}
\sum_{i=1}^N w_i = \mathbf{w}^T \cdot \mathbf{1} =1
\end{equation}
where $\mathbf{1}$ is a vector of ones.  If additionally
$\label{rown:12} \forall _i \quad w _i \geq 0$ the short sell is
excluded. Denoting $\mathbf{R}$ as a vector of expected returns of
single stocks, we see, that an expected return of a whole
portfolio is a linear combination of returns of assets in a
portfolio \[R_p =\sum_{i=1}^Nw_i\cdot R_i =\mathbf{w}^T\cdot
\mathbf{R}\] To calculate the risk of a given portfolio we
introduce a certain metric of interdependence between different
random variables. The most natural one is the statistic measure -
covariance $cov_{i,j}$, which expresses the interdependence of
variables $r_{i,t}$ and $r_{j,t}$ in all observed discrete times
$t= 1,\ldots,T$.
\begin{equation}\label{rown:14}
cov_{i,j}= \frac{1}{T} \sum_{t=1}^T(r_{i,t} - R_i)\cdot(r_{j,t} -
R_j)\Leftrightarrow \frac{1}{T}\mathbf{M}^T
\mathbf{M}=\mathbf{Cov}
\end{equation}
Now we are ready to define the variance of a portfolio as
\begin{equation}\label{rown:18}
\mathbf{\sigma_p^2 =w^T\cdot Cov \cdot w}
\end{equation}
\subsubsection{Optimization of a Portfolio} We can calculate the
return and risk of any given portfolio. Now we have to find and
choose the effective portfolios. Since it is the quadratic
programming problem, it will be done in two steps
\begin{enumerate}
\item First; the portfolio with minimal risk of all possible
portfolios will be selected (the return rate is equal to zero, ie
$R_p = 0$)
\item Secondly; we will find the minimum variance portfolio among portfolios
of arbitrary chosen return rate $(R_p=\mu)$ and then find the
efficient frontier iteratively
\end{enumerate}
\textbf{Minimal Risk Portfolio}\\
We have to find the vector of weights $\mathbf{w}$. In order to do
it we need to know perfectly the covariance matrix \footnote{This
is a very strong assumption, since as we shall see later,
covariance matrix derived from empirical data contains a high
amount of noise and statistical uncertainty.}. Let $f$ is the
function of risk, depending  of portfolio composition
\begin{equation}\label{rown:32}
f(p) = \mathbf{\sigma_p^2 = w^T\cdot Cov \cdot w}
\end{equation}
with linear constraint (\ref{rown:11})
\begin{equation}\label{rown:33}
\mathbf{w}^T \cdot \mathbf{1} =1
\end{equation}
Our task is  to minimize the function $f$ under the linear
constraint (\ref{rown:11}). This can be done in a convenient way
by using the method of \emph{Lagrange multipliers}. We get the
Lagrange function in a form:
\begin{equation}\label{rown:24}
F(\mathbf{w,\lambda}) = \mathbf{w^T \cdot Cov \cdot w
+\lambda}(\mathbf{w^T\cdot 1}-1)
\end{equation}
Standard methods of finding the minimum of a multivariate function
with a boundary condition lead to the system of $N+1$ equations
with $N+1$ unknown quantities
\begin{equation}\label{rown:25}
\left\{
\begin{array}{rcl}
  \mathbf{ 2\cdot Cov \cdot w + \lambda \cdot 1 }&=& 0 \\
  \mathbf{w^ T \cdot 1} &=& 1
\end{array}
\right.
\end{equation}
\textbf{Minimal Variance Portfolio}\\ Second task contains one
more restriction, that the expected return of a portfolio $p$ have
to obey:
\begin{equation}\label{rown:19}
\mathbf{R_p = w^T\cdot R}= \mu
\end{equation}
Then the Lagrange function reads:
\begin{equation}\label{rown:20}
F(\mathbf{w, \lambda, \gamma}) = \mathbf{w^T \cdot Cov \cdot w} +
\mathbf{\lambda} (\mathbf{w^T\cdot1}-1) +
\mathbf{\gamma}(\mathbf{w^T\cdot R}- \mu)
\end{equation}
which gives us
\begin{equation} \label{rown:21}
\left\{
\begin{array}{rcl}
\mathbf{2\cdot Cov\cdot w +\lambda\cdot 1 +\gamma \cdot R}&=&0 \\
 \mathbf{w ^T \cdot 1} &=& 1 \\
  \mathbf{w^T \cdot R} &=& \mu
\end{array}
\right.
\end{equation}
In this case we have to deal with the system of $N+2$ equations
with $N+2$ unknown quantities, which is solvable in general case.
\section{Covariance Matrix and Portfolio Construction} Covariance
Matrix plays an important role in the risk measurement and
portfolio optimization. Modern Portfolio Theories assume, that
covariances or equivalently correlations between different stocks
are perfectly known and can exactly be derived from the past data.
In practice it is quite opposite. Empirical Covariance Matrices,
built from historical data enclose such a high amount of noise,
that at first look they can be treated as random. This means, that
future risk and return of a portfolio are not well estimated and
controlled. Only after the proper denoising procedure is involved,
one can construct an efficient portfolio using Markowitz's
result.\\In this section we will briefly explain how using the RMT
one can reduce the bias of the empirical Covariance Matrix.
\subsection{Gaussian Correlated Variables}
Suppose now, that the returns from different stocks are Gaussian
random variables. The joint probability distribution function can
be then written as:
\begin{equation}\label{rown:29}
P_G(M_1,M_2,\ldots,M_N)= \frac{1}{\sqrt{(2 \pi)^N
\det\mathbf{Cov}}}\cdot \exp\left[ -\frac{1}{2}\sum_{i,j}M_i\cdot
(\mathbf{Cov}^{-1}_{ij})\cdot M_j \right]
\end{equation}
where $(\mathbf{Cov}^{-1}_{ij})$ is the element of the inverse
covariance matrix.\\It is well known result, that any set of
correlated Gaussian random variables can always be decomposed into
a linear combination of independent Gaussian random variables. The
converse is also true, since the sum of Gaussian random variables
is also a Gaussian random variable. In other words, correlated
Gaussian random variables are fully characterized by their
covariance (or correlation) matrix.\footnote{This is not true in
general case, when one needs to describe the interdependence of
non Gaussian correlated variables}
\subsubsection{Covariance
estimator:} The simplest way to construct the covariance matrix
estimator for Gaussian random variables is to deal with historical
time series of returns. The empirical covariance matrix of returns
$r_{i, t}$ can be then expressed through the equation
(\ref{rown:14})
\subsection{RMT based data filtering and denoising procedure- the
shrinkage method} For any practical use of Modern Portfolio
Theory, it would be necessary to obtain reliable estimates for
covariance matrices of real-life financial returns (based on
historical data).Thus a reliable empirical determination of a
covariance matrix turns out to be difficult. If one considers $N$
assets, the covariance matrix need to be determined from $N$ time
series of length $T \gg N$. Typically $T$ is not very large
compared to $N$ and one should expect that the determination of
the covariances is noisy. This noise cannot be removed by simply
increasing the number of independent measurement of the
investigated financial market, because economic events, that
affect the market are unique and cannot be repeated. Therefore the
structure of the matrix estimator is dominated by 'measurement'
noise. From our point of view it is interesting to compare the
properties of an empirical covariance matrix $\mathbf{Cov}$ to a
purely random matrix, well defined in the sense of Random Matrix
Theory \cite{GuhrT:2}.  Deviations from the RMT might then suggest
the presence of true information. \subsubsection{Gaussian
filtering} We will assume here that the only randomness in the
model comes from the Gaussian Probability Distribution. In order
to describe the filtering procedure we will first summarize some
well known universal properties of the random matrices.
\subsubsection{RMT predictions for behaviour of eigenvalues} Let
$M$ denotes  $N \times T$ matrix, whose entries are i.i.d. random
variables, which are normally distributed with zero mean and unit
variance. As $N, T \to \infty$ and while $Q = \frac{T}{N}$ is kept
fixed, the probability density function for the eigenvalues of the
Wishart matrix $\mathbf{\tilde{Cov}= \frac{1}{T}\tilde{M} \cdot
\tilde{M}^T}$ is given by (Mar\v{c}enko, Pastur \cite{Marce:19})
\begin{equation}\label{rown:34}
\rho(\lambda) =\frac{Q}{2\pi \sigma^2} \cdot
\frac{\sqrt{\left(\lambda_{\max} -\lambda \right) \left(\lambda -
\lambda_{\min}\right)}}{\lambda}
\end{equation}
for $\lambda$ such that $\lambda_{\min}\leq \lambda \leq
\lambda_{\max}$ where $\lambda_{\min} $and $\lambda_{\max}$
satisfy
\begin{equation}\label{rown:35}
\lambda_{\min}^{\max} =\sigma^2( 1 +\frac{1}{Q} \pm
2\sqrt{\frac{1}{Q}})
\end{equation}
\subsubsection{Standard denoising procedure and the shrinkage
method:} To remove noise we need first to compare the empirical
distribution of the eigenvalues of the covariance matrix
(\ref{rown:14}) of stocks (in our case for Warsaw Stock Exchange
shares) with theoretical prediction given by ((\ref{rown:34})
-"Wishart Fit"), based on the assumption that the covariance
matrix $\mathbf{\tilde{Cov}= \frac{1}{T}\tilde{M} \cdot
\tilde{M}^T}$ is random.
\begin{figure}[h]\label{spectrum}
\begin{center}
  \includegraphics [width=10cm]{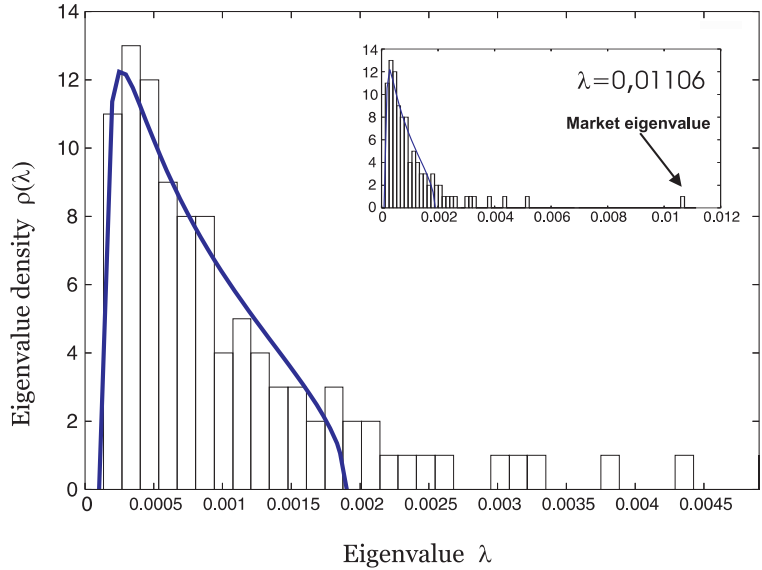}\\
  \caption{Histogram of eigenvalues for the WIG stocks from $29.01.1999$ till $17.01.2003$ with Wishart fit}\label{complete_spectrum}
\end{center}
\end{figure}
\\If we look closely at Fig. 1 we can observe, that there are
several large eigenvalues (the largest one is labeled as  the
market  one, since it consists the information about all the
stocks in the market i.e. is closely related to the WIG index),
however the greater part of the spectrum is concentrated between
$0$ and $0.002$ (i.e.   The Wishart- fit ). We believe, that
behind this  Random  part of the spectrum there exists single
eigenvalue, which carries nontrivial and useful information.
Exploiting the knowledge from Linear Algebra,we may rewrite our
covariance matrix $\mathbf{Cov}$ as:
\begin{equation}\label{rown:300}
  \mathbf{Cov = U\cdot D \cdot U^T}
\end{equation}
Here $\mathbf{D}$ is a diagonal matrix of eigenvalues  of the
original matrix $\mathbf{Cov}$ and $\mathbf{U}$ is a matrix whose
columns  are normalized eigenvectors corresponding with proper
eigenvalues. Furthermore $\mathbf{U}$ fulfills the equation:
\begin{equation}\label{rown:301}
\mathbf{U\cdot U^T = \mathrm{1} = U\cdot U^{-1}}
\end{equation}
The trace is conserved, so we write:
\begin{equation}\label{rown:302}
  \mathbf{Tr(Cov) = Tr (U\cdot D \cdot U^T)}
\end{equation}
Using the (\ref{rown:301}) and cyclic properties of the trace we
get
\begin{equation}\label{rown:303}
\mathbf{Tr (D) = Tr (Cov)}
\end{equation}
Following the fact, $\mathbf{D}$ is a diagonal matrix of
eigenvalues one can decompose its trace in the following way:
\begin{equation}\label{rown:304}
\mathbf{Tr(Cov)} =\mathbf{Tr(D)}=\sum_i \lambda_i +\sum_j
\lambda_j
\end{equation}
where $\lambda_i \in [\lambda_{\min}, \lambda_{\max}]$ is a set of
eigenvalues that are predicted by (\ref{rown:34}) $\lambda_j \in
[\lambda_1, \lambda_{\min})\cup (\lambda_{\max}, \lambda_N]$ is
set of these eigenvalues, which do not obey the RMT conditions. If
we now replace $\sum_i \lambda_i$ by one eigenvalue $ \zeta$, we
get
\begin{equation}\label{rown:305}
\zeta = \mathbf{Tr(Cov)} - \sum_j \lambda_j
\end{equation}
This results in squeezing the  Random  part of the spectrum to a
single degenerated eigenvalue. The diagonalized matrix has now
only several eigenvalues.
\subsection{Covariance Matrix Reconstruction}
Due to noise - removing procedures we know exactly the eigenvalues
of the real covariance matrix. But since we have no knowledge of
the original covariance matrix, we do not have enough knowledge of
it's eigenvectors. The familiarity with of eigenvalues is not
sufficient to find the covariance matrix. \\
After applying the denoising procedure we will reconstruct the
covariance matrix using the diagonalized matrix with some
eigenvalues shrinked and matrices of eigenvectors calculated for
 non-shrinked  covariance matrix.\\This reconstructed and unbiased
Covariance Matrix is used as an initial Covariance Matrix in
Markowitz Model described above. The new model itself is a part of
automatic investing algorithm described in the next section. The
results are presented in the last section.
\section{An Overview of the System - Automatic Investing Algorithm}
The Automatic Trading Agent is a client- server application for
managing stock portfolios without involving user interference. It
consists of three main parts: Virtual Agent, Data Collector and
User Interface. Clients running the System on their workstations
are able to monitor a stream of data (information about the state
of a portfolio) from the ATA server using their web browsers. This
part of application is controlled by the User Interface. In
addition to different standard portfolio management tools ATA
system includes several RMT - based techniques for building an
optimal portfolio with the noise effect minimized.The system is
designed not only to help a single client choose the right,
optimal portfolio with a user-defined level of risk and expected
return, but also to diminish user engagement in stock data and
information analysis. Once the strategy is fixed, client is able
to monitor the future changes in the portfolio; the rest including
portfolio optimization, data picking, sending requests and buy /
sell orders is done by a decision system - Virtual Agent.
\begin{figure}[h]
  \begin{center}
  \includegraphics [width=9cm]{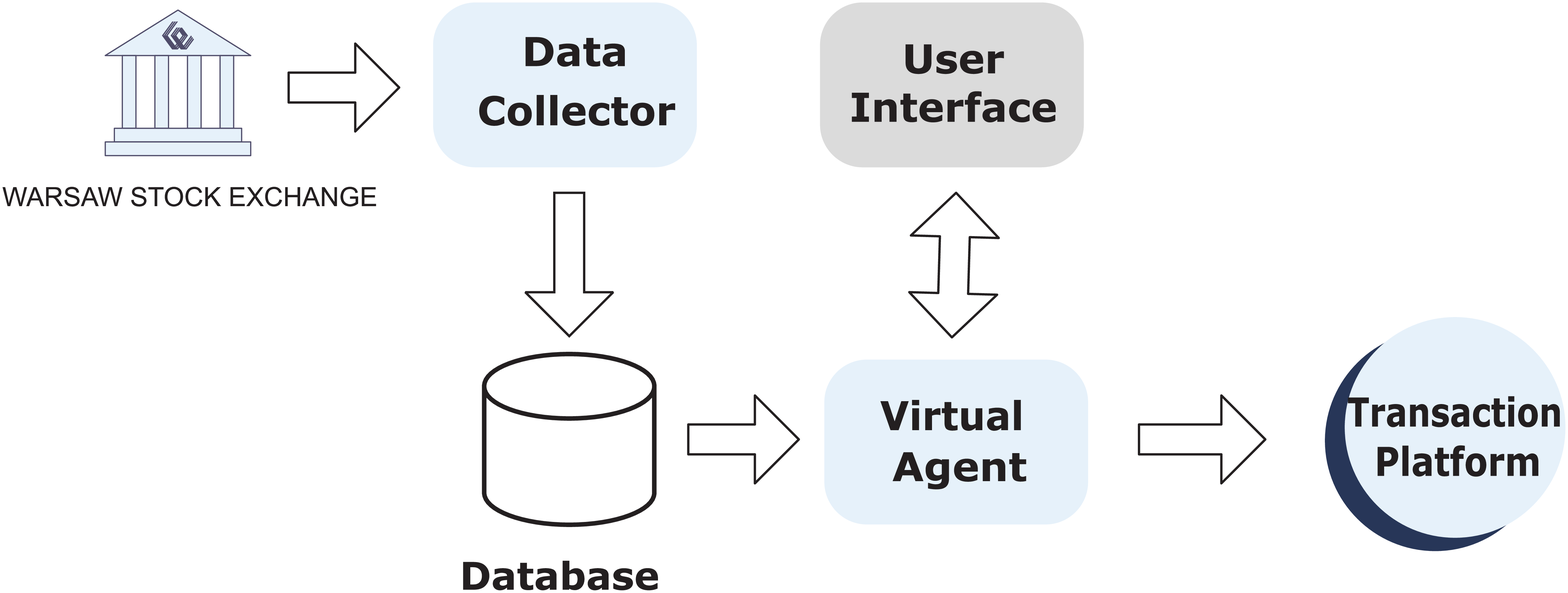}\\
  \caption{The Architecture of the System}\label{budowa}
\end{center}
\end{figure}
\subsection{Database module and Data Collector} This part of the
program is responsible for assembling and managing the stock
data.It also verifies the database in accordance with the assets
available for the transaction platform.
\subsubsection{Database}
The data are stored on the server as files with daily quotations
in a separate folder. Any company is represented by a text file,
whose name is the company's ISIN number. Each file consist of two
columns - one representing the dates and the second corresponding
daily closing prices.
\subsubsection{Data Collection}
Data collector is a separate program run by the server each
trading day, one hour after the daily quotations are closed. It
downloads the current quotation from stock  exchange data vendors
(in our case \url{http://www.parkiet.com}) and writes it down into
the database.\\ The matrix of stocks, which will be used in
further portfolio analysis, is then filled with the data from the
database.The algorithm loads all prices of securities for a
certain time window from the previously defined folder.
\subsubsection{Corrections Module}
Data are sometimes corrupted during the transfer or from
'measurement' reasons (i.e. there is no quotation for the certain
stock and the Stock Exchange is unable to state the closing
price). This result in imperfect and incomplete information and
 zeros  in initial time series. The number of files may also vary,
because it reflects
the list of assets, which are currently available for trading.\\
This part of the program watches and controls the correctness of
the files, the entries in the database and the number of files.
\subsection{Virtual Agent}
Virtual Agent is a specific
decision-making system. It's input are current and historical
stock exchange information and data form the database. On output
it generates specific requests and orders to transaction platform.
In our case it is the Stock Exchange Game structure, based on the
WARSET trading system.\\ Information conversion and data analysis
is done one hour after WSE the session is closed. All new daily
data are incorporated in the database and then optimal decision is
taken and the sell/ buy request, which will be accomplished the
next day, is sent.\\The Virtual Agent build it's resolutions on
the Effective Portfolio Theory and Random Matrix Theory.
\subsubsection{Covariance Matrix Module}
This part of the systems
offers various types of covariance matrix estimators, which are
used in solution of Markowitz's problem. The module's default
setting is the simplest Gaussian estimator (\ref{rown:14}), but
this can be modified by the user. The Covariance Matrix Module is
responsible for building a raw matrix from the data and also for
reconstructing it after the denoising procedure.
\subsubsection{Denoising and Filtering Module}
The Module controls
the diagonalization process, which uses the LU decomposition ,
i.e. calculation of eigenvalues and eigenvectors. The eigenvectors
are stored in the system and the eigenvalues are used to reduce
the degrees of freedom of the covariance matrix, as it is
predicted by RMT. Default denoising procedure is the standard one,
introduced by \cite{Bouch:23}.
\subsubsection{Portfolio
optimization}
This module is a separate program, which solves the
Markowitz's problem and finds the optimal portfolio and then sends
buy / sell order. Before any request is sent, Virtual Agent
verifies it's own decisions using several criteria. The simplest
one is to check, whether the costs of the predicted transaction
are not higher, than the realized portfolio. If they are, then
Agent sends hold request on the whole portfolio.\\ Such a
portfolio correction is usually done once a month. \footnote{The
frequency of correction, like all other key parameters can be
increased by the user}. The correction means to find once again
the portfolio with fixed level of return and risk accepted,
regarding all the new quotations since the last accomplished
correction.
\subsubsection{Corrected Portfolio:}
We have to
compare two separate portfolios: the 'old' one, which pattern is
stored on the remote transaction platform with the 'new' one,
created using the incorporated quotations. The next step is to
determine an abstract portfolio as a
result of subtractions between the examined portfolios.\\
Let $\mathbf{n}$ is the vector of weights of the new portfolio,
and $\mathbf{s}$ denotes the same vector for the old portfolio,
then the weights of a correction one are:
\begin{equation}\label{rown:43}
    \mathbf{w =n -s}
\end{equation}
If a component of $\mathbf{w}$ is $< 0$ the sell request is sent,
and obviously for $w_i >0$ system performs a buy order.$w_i = 0$
means system holds that certain asset and it's share in a
portfolio does not change.
\subsubsection{Transaction costs:}
Each
change in a portfolio is charged with brockerages (Table:
\ref{prowizje}). To compensate this effect we need to sell
slightly more individual stocks, than it arises from our analysis.
 The reverse effect has to be applied to buy request.
\subsection{Communication and Reporting Modules - User interface}
\subsubsection{Communication Module}
The communication module
allows the Virtual Agent to connect to the Game platform and place
appropriate orders. This module is a separate script, constructed
to be independent of the trading platform. This gives the
possibility to replace the Simulator used in the testing period by
the real trading platform.
\subsubsection{Reporting Module}
The
User Interface plays the role of the reporting module. It's
external part, accessible for the user is the web page
(\url{myricaria.if.uj.edu.pl}). Here the investor can follow
present information on accounts, the gains and losses figures and
the history of all changes, investment strategies and decisions
taken. The system user has also a possibility to change the key
parameters of the program, such as investment strategy (choice of
the level of risk accepted) and the frequency of portfolio
corrections.
\subsection{Implemented technologies}
\subsubsection{C\# Language}
The ATA is completely written in C\# language, chosen because of
multi-platform advantage. The programs may be written in one
environment and then run under any platform i.e. Windows and
Linux. the default environment for the ATA is linux server, but
the programming process was made under Windows, so the
multi-platform ability is a must.\\Another important advantage of
the language is the intuitive construction of mathematical
formulas and the precision of calculations far beyond the popular
C++ language, which in our case is crucial.
\subsubsection{Linux Tools}
The Data Collector is a BASH shell script, run by Cron daemon,
every fixed number of days. The script also uses Wget to
efficiently collect the data via FTP/HTTP. The AWK , SED and GREP
allow the script very easily to explore and analyze high amounts
of data.
\subsubsection{HTML, PHP, CSS}
The user interface is prepared as the website. The PHP scripts run
by the www server Apache, allow the creation of dynamic HTML
websites, where the content changes frequently. The proper view of
the website in any internet viewer is controlled by the CSS.
\section{Warsaw Stock Exchange Simulator and ATA implementation
results} Here we present the results of the whole procedure
described above. For our research we have chosen the Warsaw Stock
Exchange simulator available via the world world wide web url
\url{http://gra.onet.pl}, as a testground.
\subsection{Rules of the game}
There are several steps and rules a user must adhere and execute
to properly use the simulator. First of all, the system needs to
recognize us as its' users, possessing so called
\textsl{onet\_id}. Thus the primary step is to register oneself in
the onet system, by filling out a simple form. Using
\textsl{onet\_id} one may now log on \url{http://gra.onet.pl} to
create our first account, with $40000$ PLN as an initial sum of
money for every account. The number of accounts a single user may
open is not limited and the money can be arbitrarily invested.
Sharing more than one account
number, one is able to check different investment strategies.\\
This game act like a real stock exchange and brockerage house. We
have to start with buy order - choose financial instruments, which
we want to buy and specify their quantity and price limits. If
there are no constraints on price, then the order is realized at
any price. All quotations are delayed $20$ minutes, to give the
same chance to the players who cannot follow the quotations in
real time. All orders are cancelable, also with $20$ minutes
delay.Each user also has to pay transaction costs as in Table
\ref{prowizje}
\begin{table}[h]
  \centering
  \begin{tabular}{|c|c|}\hline
  Value of order & Height of brockerage \\ \hline \hline
  $\leq 500$ PLN & $10$ PLN \\ \hline
 $500 $--$2500$PLN & $10$PLN $+1,5\%$ over $500$ PLN \\ \hline
 $2500 $--$10000$PLN & $40$PLN $+1\%$ over $2500$ PLN \\ \hline
$\geq 10000$PLN & $115$PLN $+0,75\%$ over $10000$ PLN \\ \hline
\end{tabular}
  \caption{Costs and Commission
  (source:  \url{http://gra.onet.pl/nowa/prowizje.asp})}
  \label{prowizje}
\end{table}\\
We have constructed a certain portfolio, after our buy order is
being  accomplished. Now we need to decide, what shares we need to
buy / sell / hold to minimize the risk and maximize the return. To
win an excellent rank and high gains, one need to be involved and
follow the price changes permanently. Most of the steps one need
to execute, except the choice of the accepted level of risk, can
be done automatically by especially programmed virtual agent.
\subsection{Data Selection and Analysis}
The WIG index incorporates about $120$ stocks, which make about
$80\%$ of all assets quoted during continuous trading. From our
point of view, it is interesting to examine the connections (i.e.
correlations) between these stocks.\\ In order to conduct further
research and improve the effectiveness of our algorithm, we first
need to identify and choose a stable period in the economy.
\begin{figure}[h]
\begin{center}
  \includegraphics [width=9cm]{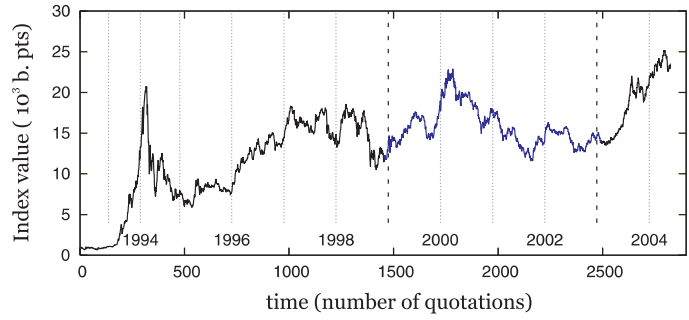}\\
  \caption{ Changes in WIG index during the period from $1991$ till $2004$ }\label{WIG}
\end{center}
\end{figure}
We have related it with the period of the lowest volatility of the
WIG index. We have started with the conversion of absolute changes
of the WIG time series $S(t)$ to the relative ones according to
\begin{equation}\label{rown:40}
G(t) = \frac{S(t+1) - S(t)}{S(t)}
\end{equation}
\begin{figure}[h]
  \begin{center}
  \includegraphics [width=9cm]{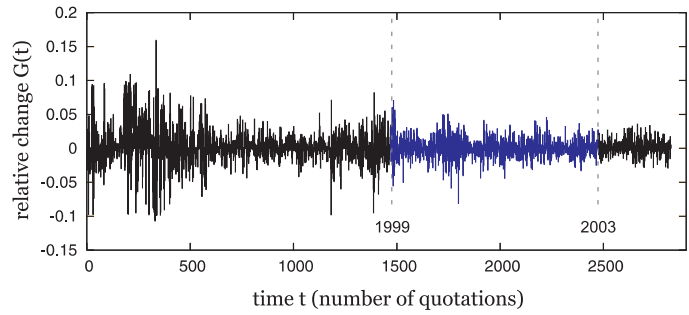}\\
  \caption{Fluctuations of relative WIG changes}\label{fluktuacje}
\end{center}
\end{figure}
Then for a fixed time window width $T= 990$ quotations, the
volatility of the time series $G(t)$ was calculated:
\begin{equation}\label{rown:41}
\sigma (t_0) = \sqrt{\frac{1}{T-1}\sum_{i=0}^T\left(G(t_0 +i)
-\overline{G(T)}\right)^2}
\end{equation}
where $\overline{G(T)}$ is the average $G(t)$ over the whole time
window $T$.This results can be presented on the diagram:
\begin{figure}[h]
\begin{center}
\includegraphics [width=9cm]{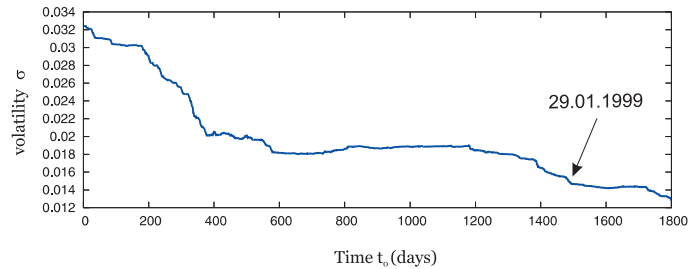}\\
  \caption{Volatility changes in time for a fixed window length}\label{990}
\end{center}
\end{figure}
It is easy to notice, that first few years of quotations are
determined by a relatively high volatility. That is why the period
from $29.01.1999$ to $17.01.2003$ was chosen in further analysis
and
tests.\\
Another problem we have encountered during the analysis of
historical data, was the incomplete information about some of
$120$ stocks, which may result in the infinities in relative
changes $G(t)$, when the lack of information was replaced by zeros
in the original $S(t)$ time series \footnote{'Zeros'appear when
one is unable to settle the price of an individual stocks, see
Ziębiec (2003) }. The separate 'zeros' were extrapolated from the
future and previous relative changes of a given time series. In
the case, if more information is lost in the way, one is unable to
predict the further prices then this stock is not very examined in
further research. For the fixed period of $990$ days we have
chosen then $100$ stocks and we have calculated the average
standard deviation of price changes $\left\langle \sigma
\right\rangle = 0,4767$ and average correlation of returns between
stocks $\left\langle corr _{ij}\right\rangle = 0,0657$
\begin{figure}[h]
  \begin{center}
  \includegraphics [width=10cm]{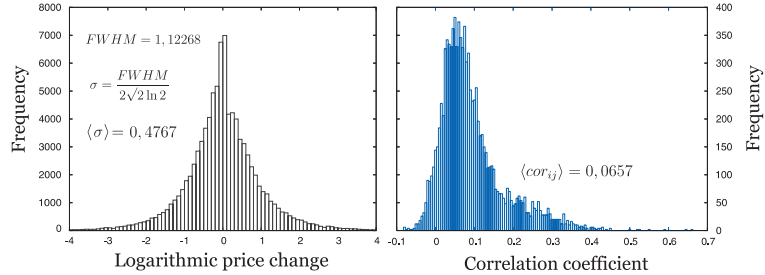}\\
  \caption{Logarithmic price changes(on the left) and correlations (on the right) for WIG companies}\label{histogramy}
\end{center}
\end{figure}
\subsection{Simulation on the historical data and it's results}
The next step in testing our system is to check how it works when
the input and output data are historical. The selected time period
was divided into parts. We have assumed, that the initial value of
a portfolio is $40000$ PLN. We have used here a time window with
variable width $T$. The analysis started with $T =139$ days. Every
day, the $T$ - dimension of the matrix $\mathbf{M}$ was increased
by one, until the final $T= 849 $ days.\\ The number of available
stocks $N = 100$ and the average number of stocks selected $45$.
Every $316$ days the correction was made. The  portfolio went to
the roof on $151$ day with $56443,61$ PLN a a result. (This is
$140\%$ of the initial value). \\The portfolio went to the floor
with $35042, 66$ PLN after  first $27$ days. \\ The result of the
investment after $ 849$ days yields $47185, 86$ PLN ,which means
the $18\%$ gains compared to the $10\%$ WIG downfall.
\begin{figure}[h]
  \begin{center}
  \includegraphics [width=9cm]{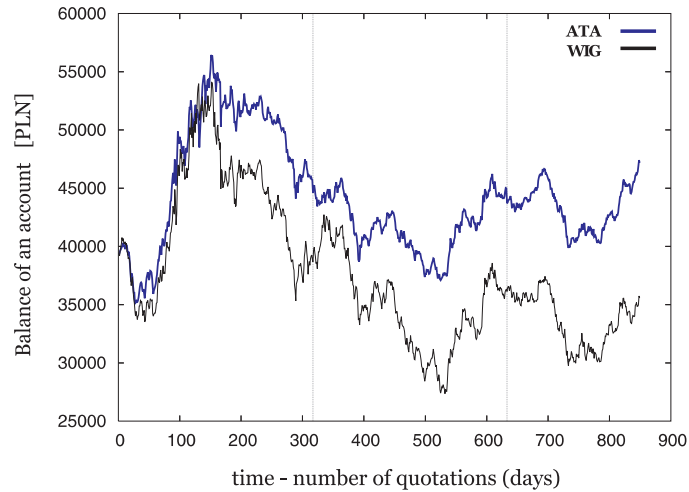}\\
  \caption{The Results of The Simulation on a Portofolio's Balance. Vertical lines indicate the
  correction}\label{ast_wig}
\end{center}
\end{figure}
\section*{Conclusions and Future Work}
\addcontentsline{toc}{chapter}{Conclusions and Future Work} The
aim of this paper was to introduce a simple RMT based mechanism,
acting like an virtual trader in a portfolio selection and
optimization process.Imposing the results from Random Matrix
Theory our program reduces the statistical noise and gives a
better estimation of future risk and return for a certain
portfolio.However, in this paper only the simplest version of the
programm was presented. An improvement of the program,  which
adopt it's decisions to the all information available will be the
part of our future work. From our point of view, an interesting
for further analysis is the hypothesis, that there exist also time
correlations between different shares. This fact might be useful
in the detection of buy/ sell  signals.

\end{document}